\documentclass[12pt,preprint]{aastex} 
\usepackage{epsf}
\usepackage[dvips]{epsfig}

\slugcomment{to appear in ApJL}
\shorttitle{The White Dwarf Deficit in Open Clusters}
\shortauthors{Fellhauer et al.}

\begin{document}

\title{The White Dwarf Deficit in Open Clusters: Dynamical Processes}

\author{M. Fellhauer, D.N.C. Lin, M. Bolte}
\affil{UCO/Lick-Observatory, University of California, Santa Cruz, CA
  95064}
\email{mike, lin, bolte @ucolick.org}
\author{S.J. Aarseth}
\affil{Institute of Astronomy, Madingley Road, Cambridge CB3 0HA, UK}
\email{sverre@ast.cam.ac.uk}
\author{K.A. Williams}
\affil{Steward Observatory, 933 N. Cherry Av., Tucson, AZ 85721}
\email{kurtis@as.arizona.edu}

\begin{abstract}
In Galactic open clusters, there is an apparent paucity of white
dwarfs compared to the number expected assuming a reasonable initial
mass function and that main-sequence stars with initial mass $< \sim
8$~M$_{\odot}$ become white dwarfs.  We suggest that this lack of
white dwarfs is due at least in part to dynamical processes.

Non-spherically symmetric mass loss during the post-main-se\-quence
evolution would lead to a few kms$^{-1}$ isotropic recoil speed for
the white dwarf remnant.  This recoil speed can cause a substantial
fraction of the white dwarfs formed in a cluster to leave the system. 

We investigate this dynamical process by carrying out high-precision
N-body simulations of intermediate-mass open clusters, where we apply
an isotropic recoil speed to the white dwarf remnants.
Our models suggest that almost all white dwarfs would be lost from the
cluster if the average recoil speed exceeds twice the velocity
dispersion of the cluster.
\end{abstract}

\keywords{open clusters and associations: general --- white dwarfs --- 
  methods: N-body simulations}

\section{Introduction}
\label{sec:intro}

Weidemann (1977) first argued that the number of white dwarf (WD)
members of the Hyades cluster was unexpectedly low.  Although
membership information was quite incomplete in 1977, reasonable
estimates for the predicted number of initial massive stars that could
have evolved into WDs suggested that approximately half the WDs were
missing.  The cooling time for the faintest Hyades WD was also less
than half the cluster age, again pointing to missing WDs, specifically
the oldest ones.  Adopting an initial mass function and assuming that
stars with initial main-sequence mass $<6$~M$_{\odot}$ evolve to
become WDs, Weidemann et al.\ (1992) quantified the Hyades WD
`deficit'.  With these assumptions, $28$ WDs are predicted to have
been formed in the cluster. After three decades of searching, only
seven WD members are known.  This deficiency of WDs in young open
clusters has been seen in the few other open clusters for which
extensive searches for WDs have been made (e.g.\ Kalirai et al.\
2001).

As has been recognized by all of the groups working in this area
through the years, there are three obvious explanations for the
missing WDs in intermediate-age clusters. First, it is possible that,
in significant conflict with stellar evolution models, the critical
initial mass above which stars explode as core-collapse supernovae is
much less than the canonical value of $6 - 8$~M$_{\odot}$.  The
presence of the lone Pleiades WD member, LB1497, with a progenitor
mass of at least $7$~M$_{\odot}$ (Claver et al.\ 2001) is not consistent
with this explanation. Second, because of their low luminosity, WDs
are difficult to detect as members of binary systems. As demonstrated
most recently by Williams (2003), with reasonable assumptions for the
binary fraction and mass-ratio distribution, a significant fraction
(20 - 50\%) of WDs are likely to be hidden in binary systems.  This is
at least a part of the explanation.  Finally, most open clusters are 
steadily losing stars due to dynamical evaporation.  The possibility
that WDs have been {\it preferentially} lost from clusters has been
investigated in various ways starting with Aarseth \& Woolf (1972),
Wielen (1974) and Pels, Oort \& Pels-Kluyver (1975) and is still a
topic of active investigation today (e.g.\ Hurley \& Shara 2003).
Although there is no unanimous consensus among the studies, it is
generally agreed that preferential evaporation of WDs from clusters is
not a significant contributor to the observed WD deficit.

Here we consider an alternative scenario, as already proposed by
Weidemann (1977), in which the paucity of WDs is due to their escape
from the cluster potential as a consequence of recoil speed attained
during the non-spherically-symmetric loss of their red giant envelope.
The typical velocity dispersion of the Galactic open clusters is
$<2$kms$^{-1}$ \ (e.g. the Hyades one-dimensional velocity dispersion
is 0.3 kms$^{-1}$; de Bruijne, Hoogerwerf \& de Zeeuw 2000).  The mass
of the main-sequence-turnoff stars is $\sim 5$~M$_{\odot}$ for a
$100$~Myr old cluster and $2$~M$_{\odot}$ for a 1 Gyr old cluster.
The typical mass of the first WD remnants is $\sim 0.8$~M$_{\odot}$
(Weidemann 2000; Claver et al.\ 2001).  The first cluster stars to
become WDs therefore lose 80\% or more of their initial mass through a
combination of stellar winds and planetary nebula ejections.  Typical
stellar wind velocities for the giant-branch phases of stars that
become the first WDs in a cluster are on the order of 10 kms$^{-1}$
(Kudritzki \& Reimers, 1978) and even the `slow', high-density
planetary nebula wind is typically expanding away from the central
star at 20 kms$^{-1}$ (e.g. Kaler \& Aller, 1974).   Thus, even a 1\%
deviation from spherical symmetry in the integrated mass loss history
would lead to a recoil speed of a few kms$^{-1}$ for the first WDs
formed in clusters.  Spruit (1998) argues that an asymmetric mass
loss fraction of the order of $10^{-3}$ during the AGB phase could
explain the rotation period distribution of WDs.  A larger asymmetric
mass loss would also induce a non negligible recoil speed.
Furthermore the author points out that such asymmetries can in
principle be observed by proper motion studies of the clumps in
interferometric images of SiO maser emission.  A similar problem has
been considered for neutron stars (Spruit \& Phinney 1998).  The
recoil speed of many pulsars is observed to be greater than 100
kms$^{-1}$ (Hansen \& Phinney 1997).  These large speeds are probably
induced by a non-spherically-symmetric supernova explosion.  
Monte Carlo simulations of supernova explosions in primordial binaries 
show that large
recoil speed leads to a substantial loss of the neutron star remnants
even in rich and strongly bound globular clusters (Pfahl et al.\ 2002).

We consider a much less volatile situation in which the WDs are formed
in an open cluster environment with a few kms$^{-1}$ recoil speed.  In
\S2, we briefly describe our numerical method and the formulation of
the problem.  The results of our numerical computation is presented in
\S3 and we discuss their implications in \S4.

\section{Numerical scheme and model parameters}
\label{sec:setup}

We perform N-body simulations using the direct N-body code {\sc
 Nbody6} (Aarseth 1999).  This numerical scheme enables us to follow
the orbits of the stars in an open cluster with high accuracy.  It is
a direct summation code with block time-steps, i.e.\ the time-steps
are quantized to powers of $2$ (Makino 1991) and an Ahmad--Cohen
neighbour scheme, which splits the force polynomial of a particle
into an irregular part due to the neighbouring particles and a
regular part of the more distant particles (Ahmad \& Cohen 1973).  It
has a Hermite integrator which is a fourth-order predictor--corrector
scheme with coordinate truncation error proportional to $\Delta
t^{5}$ (Makino \& Hut 1988).  Binaries and close encounters between
two stars are treated by the Kustaanheimo--Stiefel (KS) regularization  
method (Kustaanheimo \& Stiefel 1965).  Close encounters between
single stars and binaries, binaries and binaries or even higher
multiplicity of close particles are studied by a special method, known 
as chain regularization (Mikkola \& Aarseth 1993, Mikkola 1997).  The
code is also able to treat primordial binaries and has a scheme to
implement the stellar evolution of the stars in the cluster (Hurley,
Pols \& Tout, 2000). 

For our open cluster initial models we chose a multi-mass King model
(King 1966, Michie \& Bodenheimer 1963) with concentration parameter
$W_{0}=5$.  We perform 
simulations with $N_{\rm tot}= 2000$ and $10\,000$ particles.  The
tidal field is adjusted to a Galactic central distance of $10$~kpc.
We perform simulations with different initial binary fractions
($f_{\rm b} = 0$, $0.2$, $0.4$ and $0.8$).  The initial mass function
(IMF) is taken from Kroupa, Tout \& Gilmore (1993). The binary
population contains only hard binaries.  Most soft binaries would be
disrupted by the intra-cluster forces before the first WD is formed.
For primordial binaries the total mass of the binary was chosen from
the Kroupa-IMF, which was not corrected for the effect of binaries, 
and the component masses were then assigned according to a uniform
mass-ratio distribution.  The orbital separation was taken from the
log-normal distribution of Eggleton, Fitchett \& Tout (1989) with a
maximum of $100$~AU and the orbital eccentricity was taken from a
thermal distribution (Heggie 1975).  The properties of the two open
cluster models can be found in Table~\ref{tab:prop}.

Our low-mass model resembles an open cluster like the Hyades, which is
$7 \cdot 10^{8}$ years old (Perryman et al.\ 1998), has a tidal radius
of about $9$ - $14$~pc and currently numbers $\sim 400$ stars.  Most
likely it had $1000$ - $2500$ stars initially.

As described above, we include a recoil speed if a WD is formed.  If
the ejection of the planetary nebula which leaves a  WD behind is
anisotropic at a level of only a few percent, this mass loss would
result in a recoil velocity of the new WD in the order of a few
kms$^{-1}$.  Therefore we perform simulations with kick velocities
randomly chosen from a Maxwellian with mean velocity $v_{\rm kick} =
1$, $2$ or $5$~kms$^{-1}$.  The direction of the  kick has an
isotropic distribution.

\section{The results of numerical simulations}
\label{sec:results}

In the simulation without a kick
velocity almost all WDs remained in the system.  They also tend to be
concentrated in the cluster cores.  This result can be explained by
the relative high mass of the WD progenitors (Hurley \& Shara 2003).
N-body systems like open clusters preferentially lose low-mass stars
and some of the very high-mass stars due to multiple close encounters
in the 
centre of the system.  Stars with masses similar to WDs' progenitors
would be overrepresented in the cluster cores.  After they lose 90\%
of their initial mass the WD remnants begin to diffuse to the outer
regions of  the cluster.  However, the diffusion time scale is longer
than the two-body relaxation time scale.  Consequently, these WD
remnants remain relatively centrally concentrated. 

However, when we include a kick acquired during evolution to the
WD phase, some WDs gain enough velocity to leave the system.  In the
low-mass systems ($N=2000$) a mean kick velocity of $2$~kms$^{-1}$
depletes the number of WDs significantly.  In the case of high mass
systems ($N=10\,000$) a higher mean kick velocity of about
$5$~kms$^{-1}$ is needed to deplete the cluster of a significant
fraction of its WDs.

The results show that WDs are depleted significantly if the mean kick 
velocity exceeds twice the velocity dispersion of the open cluster.
This may be understood in terms of the escape velocity which is
also approximately twice the velocity dispersion in these systems.
Hence, if the kick velocity is roughly equal to the internal velocity
dispersion, the clusters lose a noticeable amount of WDs.

The binary fraction plays only a secondary role.  With the caveat that
our sample of simulations is small (we would need many random
realisations of one set of parameters to reduce the error bars), there 
is no significant difference between the simulations with and without
initial binaries.  In principle hard binaries with high orbital
velocities together with low kick velocities could inhibit WDs from
leaving the cluster, but mostly the binaries leave the system as a
whole.  If we take into account that the progenitor of the WD is more
massive than its companion which remains on the main sequence, the
recoil has a comparable impact on the binary system as a whole.  Again
the escape velocity plays an important role here.  The dividing line
between hard and soft binaries is defined as the orbital velocity of
the binary being equal to the escape velocity from the cluster.
Therefore only kick velocities higher than the escape velocity of the
system are able to break up such binaries. 

\begin{figure}
  \begin{center}
%%    \plotone{f1.eps}
    \epsfxsize=12cm \epsfysize=12cm \epsffile{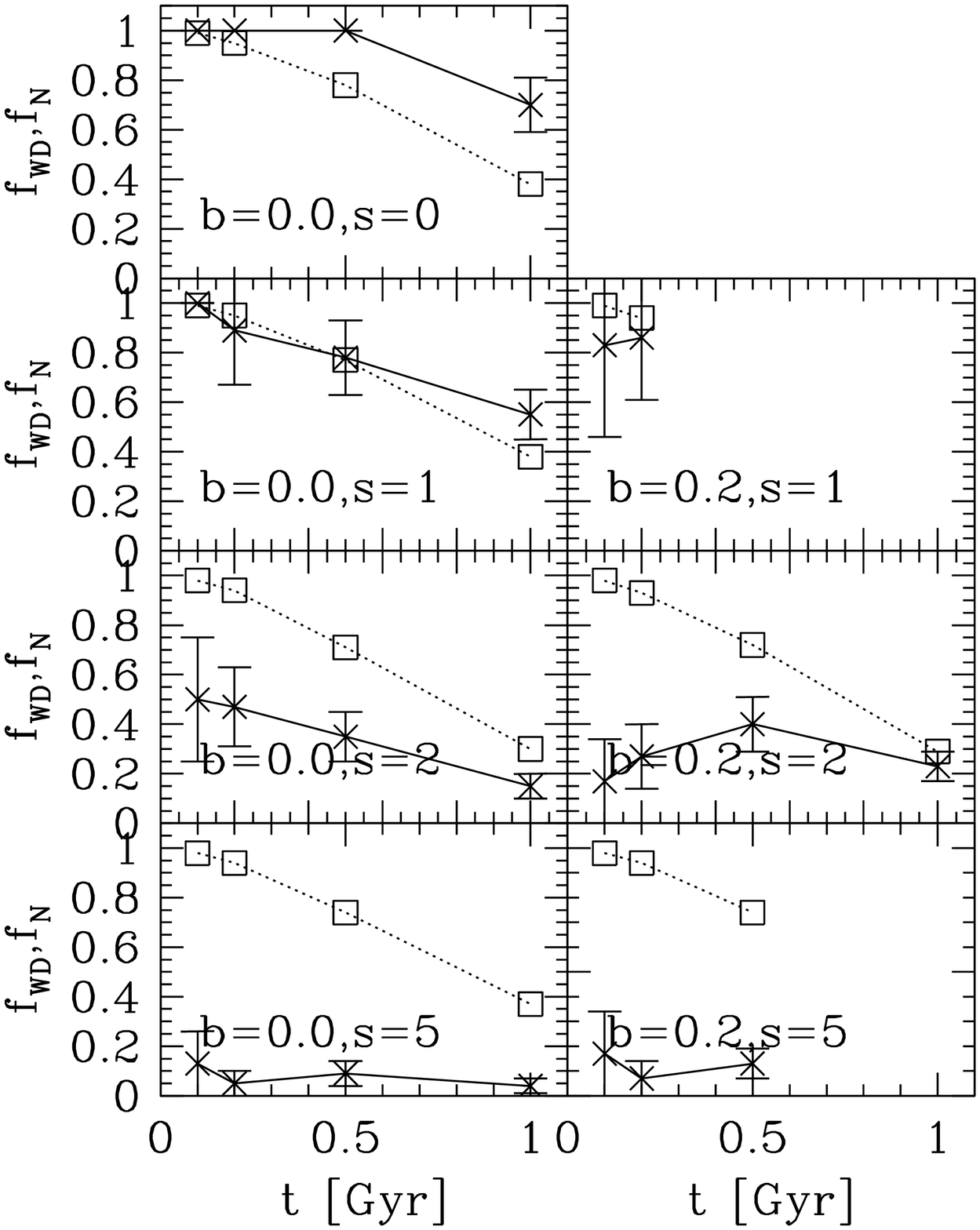}
    \caption{Time evolution of the $N=2000$ calculations. Left:
    simulations with no initial binaries; Right: simulations with an
    initial binary fraction (b) of 0.2.  From top to bottom the
    simulations with no kick velocity (s) and with mean kick velocity
    of $1.0$, $2.0$ and $5.0$~kms$^{-1}$ are shown. Crosses with
    error bars and solid lines show the fraction of WDs remaining in
    the system relative to the number of WDs produced in the system
    (f$_{\rm WD}$).  Open squares and dotted lines show the number of
    stars left in the system divided by the initial number of stars
    (f$_{\rm N}$).} 
    \label{fig:2000}
  \end{center}
\end{figure}

\begin{figure}
  \begin{center}
%%    \plotone{f2.eps}
    \epsfxsize=12cm \epsfysize=12cm \epsffile{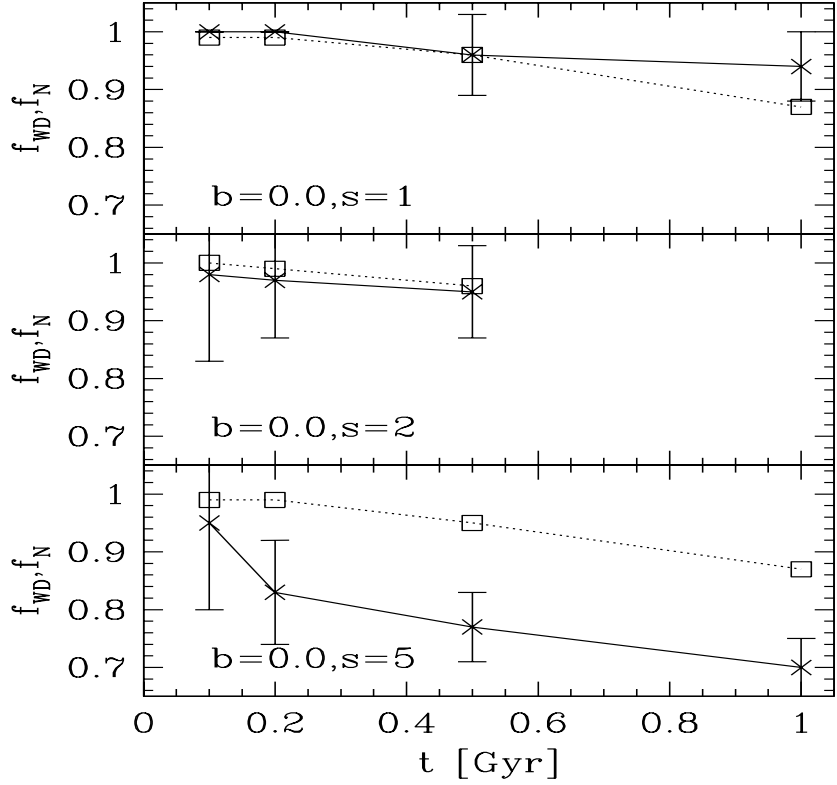}
    \caption{Time evolution of the $N=10\,000$ calculations with no
    initial binaries (b).  From top to bottom the simulations with
    mean kick velocity (s) of $1.0$, $2.0$ and $5.0$~kms$^{-1}$ are
    shown.  Crosses with error bars and solid lines show the fraction
    of WDs remaining in the system relative to the number of WDs
    produced in the system (f$_{\rm WD}$).  Open squares and dotted
    lines show the number of stars left in the system divided by the
    initial number of stars (f$_{\rm N}$).}   
    \label{fig:10000}
  \end{center}
\end{figure}

Table~\ref{tab:100myr} shows the results of our simulations taken at
$100$~Myr.  Some WDs have formed and this is also approximately one
relaxation time of the clusters.  To show the time evolution of these
results we followed certain calculations to $500$~Myr and some up to
1 Gyr.  This time evolution is shown in Fig.~\ref{fig:2000} for the
$N=2000$ cases and in Fig.~\ref{fig:10000} for the $N=10\,000$ cases.
In each plot two lines are shown.  The dotted line is the evolution of
the total number of stars in the system.  The solid line shows how
many of the produced WDs are still in the system.  If there is no kick
velocity the line of the WD fraction is above the line of all stars as
expected.  Including the kick velocity and increasing it moves the
line of the WDs downwards and as soon as the mean kick velocity
exceeds the escape velocity of the cluster the fraction of WDs is
significantly lower than the fraction of stars remaining in the
system.  This means there is a significant WD deficit in these
systems.

The remaining WDs are mainly in binaries and preferentially close to
the centre of the cluster in the simulations with low kick velocities,
and more likely to be found as single stars in calculations with high
kick velocity. 

\section{Summary and Discussion}
\label{sec:discus}

With our direct N-body simulations we have shown that if the rapid
mass loss involved in the formation of a WD progenitor is asymmetrical 
at a level of one or a few percent, the newly-formed WD suffers a
recoil kick which is able to deplete an open cluster of almost all of
its WDs.  From energy arguments, this recoil velocity is at least of
the order of a few kms$^{-1}$.  But this is already enough to exceed
the escape velocity, i.e.\ to deplete a low mass open cluster of almost
all its WDs.  More massive open clusters are able to retain WDs within
the cluster but still show a significant depletion of WDs. 

The best studied intermediate-age open clusters in the Milky Way
exhibit a deficit of WDs, even after correcting for WDs hidden in
binaries.  Our results are in agreement with the data compilation of von
Hippel (1998).  The open clusters studied by von Hippel exhibit a
steep IMF slope of $-2.35$ to $-3$.  We suggest that this deficiency
can be explained via the combined process of small kicks imparted
during the mass loss history and subsequent evaporation from the
cluster. These processes act preferentially to deplete clusters of the
first WDs to form and could mimic a steep IMF-slope.  These stars
suffer the largest amount of mass loss and statistically have the
largest kicks and have spent the longest amount of time (among the WD
population) as relatively low mass stars. This preferential loss of
the first-formed WDs may need to be accounted for when using WD
cooling times to estimate cluster ages or when tracing cluster WDs
back to the main-sequence and attempting to determine the critical 
main-sequence mass at which WDs first begin to form.

\acknowledgments
This work is partially supported (M.F., D.L.) by NASA through
NAG5-12151. M.B. and K.W. are happy to acknowledge support from the
National Science Foundation grant AST-0307492.  We also thank the
referee, Ted von Hippel, for helpful comments.

\clearpage

\clearpage

\clearpage

\begin{table}
  \begin{center}
    \caption{Properties of the open cluster models. From top to
      bottom: total mass, crossing time, relaxation time, tidal radius,
    half mass radius, core radius, velocity dispersion.}
    \label{tab:prop}
    \begin{tabular}[h!]{ccrr}
      \tableline \tableline
      $N$             &               & $2000$   & $10000$ \\
      \tableline
      $M_{\rm tot}$   & [M$_{\odot}$] & 1317.1   & 6668.1  \\
      $T_{\rm cr}$    & [Myr]         &    6.2   &    2.7  \\
      $T_{\rm relax}$ & [Myr]         &  180.7   &  109.1  \\
      $r_{\rm tidal}$ & [pc]          &   15.6   &   26.7  \\
      $r_{\rm h}$     & [pc]          &    2.5   &    2.4  \\
      $r_{\rm core}$  & [pc]          &    0.9   &    1.0  \\
      $\sigma$        & [kms$^{-1}$]  &    0.8   &    1.8  \\
      \tableline
    \end{tabular}
  \end{center}
\end{table}

\clearpage

\begin{table}
  \begin{center}
    \caption{Results of our simulations after $100$~Myr of evolution.
    The columns are the number of stars initially, the initial binary
    fraction, mean velocity of the kick in kms$^{-1}$, the number of
    WDs formed in the cluster, the number of WDs remaining in the
    cluster, the fraction of WDs remaining and the fraction of total
    stars remaining in the cluster.  Because of the small numbers in
    the $N=2000$ simulations, we performed several runs with the same
    parameters but different random realisations.  The absolute
    numbers given in the table represent the first run of each set,
    the fractions are the statistical mean out of all simulations for
    a certain parameter set.  The last column gives the number of
    random realisations performed with this parameter setting.} 
    \label{tab:100myr}
    \begin{tabular}[h!]{rrrrrrrr}
      \tableline \tableline
      $N$ & $f_{\rm b}$ & $v_{\rm kick}$ & WD$_{\rm tot}$ & WD$_{\rm
      r}$ & $f_{\rm WD}$ & $f_{\rm N}$ & \#run \\ 
      \tableline
      2000 & 0.0 & 0 & 8 & 8 & 1.00 & 0.99 & 1 \\
      2000 & 0.0 & 1 & 8 & 8 & 1.00 & 0.99 & 1 \\
      2000 & 0.0 & 2 & 8 & 4 & 0.50 & 0.98 & 1 \\
      2000 & 0.0 & 5 & 8 & 1 & 0.13 & 0.98 & 1 \\
      2000 & 0.2 & 1 & 6 & 5 & 0.83 & 0.98 & 3 \\
      2000 & 0.2 & 2 & 6 & 1 & 0.17 & 0.98 & 1 \\
      2000 & 0.2 & 5 & 6 & 1 & 0.09 & 0.98 & 2 \\
      2000 & 0.4 & 1 & 5 & 5 & 1.00 & 0.98 & 3 \\
      2000 & 0.4 & 2 & 5 & 4 & 0.80 & 0.98 & 3 \\
      2000 & 0.4 & 5 & 5 & 1 & 0.20 & 0.98 & 1 \\
      2000 & 0.8 & 1 & 3 & 3 & 1.00 & 0.98 & 1 \\
      2000 & 0.8 & 2 & 3 & 3 & 1.00 & 0.99 & 1 \\
      2000 & 0.8 & 5 & 1 & 0 & 0.00 & 0.97 & 1 \\
      10000 & 0.0 & 1 & 43 & 43 & 1.00 & 1.00 & 2 \\
      10000 & 0.0 & 2 & 43 & 42 & 0.98 & 1.00 & 1 \\
      10000 & 0.0 & 5 & 43 & 38 & 0.88 & 0.99 & 2 \\
      10000 & 0.2 & 2 & 39 & 38 & 0.97 & 0.99 & 1 \\
      10000 & 0.2 & 5 & 39 & 37 & 0.95 & 0.99 & 1 \\
      10000 & 0.4 & 2 & 27 & 27 & 1.00 & 0.99 & 1 \\
      \tableline
    \end{tabular}
  \end{center}
\end{table}


\begin{thebibliography}{}
\bibitem[]{aar72} Aarseth S.J., \& Woolf N.J. 1972, Astrophys.Lett.,
  12, 159
\bibitem[]{aar99} Aarseth S.J. 1999, \pasp, 111, 1333
\bibitem[]{ahm73} Ahmad A., \& Cohen L. 1973, J. Comp. Phys., 12, 389 
\bibitem[]{bru00} de Bruijne J.H.J., Hoogerwerf R., \& de Zeeuw
  P.T. 2000, A\&A, 367, 111 
\bibitem[]{cla01} Claver C.F., Liebert J, Bergeron P., \& Koester
  D. 2001, \apj, 563, 987 
\bibitem[]{egg89} Eggleton P.P., Fitchett M., \& Tout C.A. 1989,
  \apj, 347, 998 
\bibitem[]{han97} Hansen B.M.S., \& Phinney E.S. 1997, \mnras, 291, 569
\bibitem[]{heg75} Heggie D.C. 1975, \mnras, 173, 729
\bibitem[]{hur00} Hurley J.R., Pols O.R., \& Tout C.A. 2000, \mnras,
  315, 543 
\bibitem[]{hur03} Hurley J.R., \& Shara M.M. 2003, \apj 589, 179
\bibitem[]{kal74} Kaler J.B., \& Aller L.H. 1974, PASP, 86, 635
\bibitem[]{kal01} Kalirai J.S., Ventura P., Richer H.B., Fahlman G.G.,
  Durrell P.R., D'Antona F., \& Marconi G. 2001, \aj, 122, 3239
\bibitem[]{kus65} Kustaanheimo P., \& Stiefel E. 1965, J. Reine
  Angewandte Mathematik, 218, 204
\bibitem[]{kin66} King I. 1966, \aj, 71, 64
\bibitem[]{kro93} Kroupa P., Tout C.A., \& Gilmore G. 1993, \mnras,
  262, 545 
\bibitem[]{kud78} Kudritzki R.P., \& Reimers D. 1978, A\&A, 70, 227
\bibitem[]{mak88} Makino J., \& Hut P. 1988, \apjs, 68, 833
\bibitem[]{mak91} Makino J. 1991, PASJ, 43, 141
\bibitem[]{mic63} Michie R.W., \& Bodenheimer P.H. 1963, \mnras, 126,
  269 
\bibitem[]{mik93} Mikkola S., \& Aarseth S.J. 1993,
  Celes.Mech.Dyn.Astron., 57, 439 
\bibitem[]{mik97} Mikkola S. 1997, in Visual Double Stars: Formation,
  Dynamics and Evolutionary Tracks, eds.\ Docobo J.A. et al.\
\bibitem[]{pel75} Pels G., Oort J.H., \& Pels-Kluyver H.A. 1975,
  A\&A, 43, 423 
\bibitem[]{per98} Perryman, M.A.C., Brown A.G.A., Lebreton Y., Gomez
  A., Turon C., de Strobel G.C., Mermilliod J.C., Robichon N.,
  Kovalevsky J., \& Crifo, F. 1998, A\&A, 331, 81
\bibitem[]{pfa02} Pfahl E., Rappaport S., \& Podsiadlowski P. 2002,
  \apj, 573, 283
\bibitem[]{spr98a} Spruit H.C. 1998, A\&A, 333, 603
\bibitem[]{spr98b} Spruit H.C., \& Phinney E.S. 1998 Nature, 393, 139 
\bibitem[]{hip98} von Hippel T. 1998, \aj, 115, 1536
\bibitem[]{wei77} Weidemann V. 1977, A\&A, 59, 411
\bibitem[]{wei92} Weidemann V., Jordan S., Iben I., \& Casertano
  S. 1992, \aj, 104, 1876
\bibitem[]{wei00} Weidemann V. 2000, A\&A, 363, 647
\bibitem[]{wie74} Wielen R. 1974, in Stars and the Milky Way System,
  ed.\ Mavridis L.N. (Springer Verlag), 326
\bibitem[]{wil03} Williams K.A. 2003, PhD-thesis, University of
  California Santa Cruz
\end{thebibliography}
\end{document}